\documentclass[aps,showpacs,preprintnumbers,amsmath, amssymb]{revtex4}

\usepackage{float}
\usepackage{graphics,epsfig}
\usepackage{graphicx}
\usepackage{dcolumn}
\usepackage{bm}

\begin{document}
\baselineskip=0.8 cm
\title{{\bf Analytical study on holographic superfluid in AdS soliton background}}

\author{Chuyu Lai$^{1}$, Qiyuan Pan$^{1,2}$\footnote{panqiyuan@126.com}, Jiliang Jing$^{1}$\footnote{jljing@hunnu.edu.cn}
and Yongjiu Wang$^{1}$\footnote{wyj@hunnu.edu.cn}}
\affiliation{$^{1}$ Department of Physics, Key Laboratory of Low
Dimensional Quantum Structures and Quantum Control of Ministry of
Education, and Synergetic Innovation Center for Quantum Effects and
Applications, Hunan Normal University, Changsha, Hunan 410081,
China} \affiliation{$^{2}$ Instituto de F\'{\i}sica, Universidade de
S\~{a}o Paulo, CP 66318, S\~{a}o Paulo 05315-970, Brazil}

\vspace*{0.2cm}
\begin{abstract}
\baselineskip=0.6 cm
\begin{center}
{\bf Abstract}
\end{center}

We analytically study the holographic superfluid phase transition in
the AdS soliton background by using the variational method for the
Sturm-Liouville eigenvalue problem. By investigating the holographic
s-wave and p-wave superfluid models in the probe limit, we observe
that the spatial component of the gauge field will hinder the phase
transition. Moreover, we note that, different from the AdS black
hole spacetime, in the AdS soliton background the holographic
superfluid phase transition always belongs to the second order and
the critical exponent of the system takes the mean-field value in
both s-wave and p-wave models. Our analytical results are found to
be in good agreement with the numerical findings.

\end{abstract}
\maketitle
\newpage
\vspace*{0.2cm}

\section{Introduction}

As we know, the phenomenology of conventional superconductors is
extremely well explained by Bardeen-Cooper-Schrieffer (BCS) theory
\cite{BCS} and its extensions \cite{Parks}. However, these theories
fail to describe the core mechanism governing the high-temperature
superconductor systems which is one of the unsolved mysteries in
modern condensed matter physics. Interestingly, the anti-de
Sitter/conformal field theories (AdS/CFT) correspondence
\cite{Maldacena,Witten,Gubser1998}, which can map strongly coupled
non-gravitational physics to a weakly coupled perturbative
gravitational problem, might provide some meaningful theoretical
insights to understand the physics of high $T_{c}$ superconductors
from the gravitational dual
\cite{HartnollRev,HerzogRev,HorowitzRev,CaiRev}. The main idea is
that the spontaneous $U(1)$ symmetry breaking by bulk black holes
can be used to construct gravitational duals of the transition from
normal state to superconducting state in the boundary theory, which
exhibits the behavior of the superconductor
\cite{GubserPRD78,HartnollPRL101}. In additional to the bulk AdS
black hole spacetime, it was found that a holographic model can be
constructed in the bulk AdS soliton background to describe the
insulator and superconductor phase transition
\cite{Nishioka-Ryu-Takayanagi}.

In general, the studies on the gravitational dual models of the
superconductorlike transition focus on the vanishing spatial
components of the $U(1)$ gauge field on the AdS boundary.
Considering that the supercurrent in superconducting materials is a
well studied phenomenon in condensed matter systems, the authors of
Refs. \cite{BasuMukherjeeShieh,HerzogKovtunSon} constructed a
holographic superfluid solution by performing a deformation of the
superconducting black hole, i.e., turning on a spatial component of
the gauge field that only depends on the radial coordinate. It was
found that the second-order superfluid phase transition can change
to the first order when the velocity of the superfluid component
increases relative to the normal component. Interestingly, the
holographic superfluid phase transition remains second order for all
allowed fractions of superfluid density in the strongly-backreacted
regime at low charge $q$ \cite{SonnerWithers}. However, in the case
of the fixed supercurrent, the superfluid phase transition is always
of the first order for any nonzero supercurrent
\cite{AreanJHEP2010,Zeng2011,Zeng2013}. In Ref. \cite{Arean2010},
the effect of the scalar field mass on the superfluid phase
transition was investigated and it was observed that the Cave of
Winds exists for some special mass in the superfluid model. In order
to explore the effect of the vector field on the superfluid phase
transition, a holographic p-wave superfluid model in the AdS black
holes coupled to a Maxwell complex vector field was introduced
\cite{PWavSuperfluidA,LifshitzSuperfluid} and it was revealed that
the translating superfluid velocity from second order to first order
increases with the increase of the mass squared of the vector field.
On the other hand, from the perspective of the QNM analysis, the
question of stability of holographic superfluids with finite
superfluid velocity was revisited and it was suggested that there
might exist a spatially modulated phase slightly beyond the critical
temperature \cite{Amado2013,Amado2014}.

The aforementioned works on the holographic superfluid models
concentrated on the AdS black hole configuration. More recently, the
authors of Refs. \cite{Peng2012,KuangLiuWang} extended the
investigation to the soliton spacetime and investigated numerically
the holographic s-wave superfluid model in the AdS soliton
background. It was found that, in the probe limit, the first-order
phase transition cannot be brought by introducing the spatial
component of the vector potential of the gauge field in the AdS
soliton background, which is different from the black hole spacetime
\cite{KuangLiuWang}. In order to back up numerical results and
further reveal the properties of the holographic superfluid model in
the probe limit, in this work we will use the analytical
Sturm-Liouville (S-L) method, which was first proposed in
\cite{Siopsis,SiopsisB} and later generalized to study holographic
insulator/superconductor phase transition in \cite{Cai-Li-Zhang}, to
analytically investigate the holographic s-wave superfluid model in
the AdS soliton background. Considering that the increasing interest
in study of the Maxwell complex vector field model
\cite{CaiPWave-1,CaiPWave-2,CaiPWave-3,CaiPWave-4,WuPWave-1,CaiPWave-5,
WuPWave-2,CaiPWave-6,ZPJ2015,CSJHEP2015,Nie2015,Rogatko2015}, we
will also extend the investigation to the holographic p-wave
superfluid model in the AdS soliton background, which has not been
constructed as far as we know. Besides to be used to check numerical
computation, the analytical study can clearly disclose some general
features for the effects of the spatial component of the gauge field
on the holographic superfluid model in the AdS soliton background.

The structure of this work is as follows. In Sec. II we will
investigate the holographic s-wave superfluid model in the AdS
soliton background. In particular, we calculate the critical
chemical potential of the system as well as the relations of
condensed values of operators and the charge density with respect to
$(\mu-\mu_{c})$. In Sec. III we extend the discussion to the p-wave
case which has not been constructed as far as we know. We will
conclude in the last section with our main results.

\section{Holographic s-wave superfluid model}

We start with the five-dimensional Schwarzschild-AdS soliton in the
form
\begin{eqnarray}\label{SchSoliton}
ds^2=-r^2dt^2+\frac{dr^2}{f\left(r\right)}+f\left(r\right)d\varphi^2+r^2(dx^2+dy^2),
\end{eqnarray}
where $f(r)=r^2(1-r_{s}^{4}/r^{4})$ with the tip of the soliton
$r_{s}$ which is a conical singularity in this solution. We can
remove the singularity by imposing a period $\beta=\pi/r_{s}$ for
the coordinate $\varphi$. As a matter of fact, this soliton can be
obtained from a five-dimensional AdS Schwarzschild black hole by
making use of two Wick rotations.

In order to construct the holographic s-wave model of superfluidity
in the AdS soliton background, we consider a Maxwell field and a
charged complex scalar field coupled via the action
\begin{eqnarray}\label{SWaveAction}
S=\int
d^5x\sqrt{-g}\left(-\frac{1}{4}F_{\mu\nu}F^{\mu\nu}-|\nabla_\mu\psi-iqA_\mu\psi|^2-m^2|\psi|^2\right),
\end{eqnarray}
where $q$ and $m$ represent the charge and mass of the scalar field
$\psi$ respectively. Taking the ansatz of the matter fields as
\begin{eqnarray}\label{Ansatz}
\psi=\psi(r),~~A_\mu dx^{\mu}=A_t(r)dt+A_{\varphi}(r)d\varphi,
\end{eqnarray}
where both a time component $A_t$ and a spatial component
$A_{\varphi}$ of the vector potential have been introduced in order
to consider the possibility of DC supercurrent, we can get the
equations of motion in the probe limit
\begin{eqnarray}\label{EqMotionr}
&&\psi^{\prime\prime}+\left(\frac{3}{r}+\frac{f^\prime}{f}\right)\psi^{\prime}
-\frac{1}{f}\left(m^2+\frac{q^2A^2_\varphi}{f}-\frac{q^2A_t^2}{r^2}\right)\psi=0,\nonumber \\
&&A_t^{\prime\prime}+\left(\frac{1}{r}+\frac{f^\prime}{f}\right)A_t^{\prime}-\frac{2q^2\psi^2}{f}A_t=0,\nonumber \\
&&A_\varphi^{\prime\prime}+\frac{3}{r}A_\varphi^{\prime}-\frac{2q^2\psi^2}{f}A_\varphi=0,
\end{eqnarray}
where the prime denotes the derivative with respect to $r$. From the
equation of motion for $\psi$, we can obtain the effective mass of
the scalar field
\begin{eqnarray}\label{EffectiveMass}
m_{eff}^{2}=m^2+\frac{q^2A^2_\varphi}{f}-\frac{q^2A_t^2}{r^2},
\end{eqnarray}
which implies that the increasing $m^2$ and $A_\varphi$ or
decreasing $A_t$ will hinder the s-wave superfluid phase transition.
We will get the consistent result in the following calculation.

In order to solve above equations, we have to impose the appropriate
boundary conditions at the tip $r=r_{s}$ and the boundary
$r\rightarrow\infty$. At the tip $r=r_{s}$, the fields behave as
\begin{eqnarray}\label{TipCondition}
&&\psi=\tilde{\psi}_{0}+\tilde{\psi}_{1}(r-r_{s})+\tilde{\psi}_{2}(r-r_{s})^{2}+\cdots\,, \nonumber \\
&&A_{t}=\tilde{A}_{t0}+\tilde{A}_{t1}(r-r_{s})+\tilde{A}_{t2}(r-r_{s})^{2}+\cdots\,, \nonumber \\
&&A_{\varphi}=\tilde{A}_{\varphi1}(r-r_{s})+\tilde{A}_{\varphi2}(r-r_{s})^{2}+\cdots,
\end{eqnarray}
where $\tilde{\psi}_{i}$, $\tilde{A}_{ti}$ and $\tilde{A}_{\varphi
i}$ ($i=0,1,2,\cdots$ and $\tilde{A}_{\varphi 0}=0$) are the
integration constants, and we have imposed the Neumann-like boundary
conditions to render the physical quantities finite
\cite{Nishioka-Ryu-Takayanagi}. Obviously, we can find a constant
nonzero gauge field $A_{t}(r_{s})$ at $r=r_{s}$, which is in strong
contrast to that of the holographic superfluid model in the AdS
black hole background where $A_{t}(r_{+})=0$ at the horizon
\cite{BasuMukherjeeShieh,HerzogKovtunSon,KuangLiuWang}.

At the asymptotic AdS boundary $r\rightarrow\infty$, we have
asymptotic behaviors
\begin{eqnarray}\label{InfinityCondition}
&&\psi=\frac{\psi_{-}}{r^{\Delta_-}}+\frac{\psi_{+}}{r^{\Delta_+}},~~A_t=\mu-\frac{\rho}{r^2},
~~A_\varphi=S_\varphi-\frac{J_\varphi}{r^2},
\end{eqnarray}
where $\Delta_\pm=2\pm\sqrt{4+m^2}$ is the conformal dimension of
the scalar operator dual to the bulk scalar field, $\mu$ and
$S_\varphi$ are the chemical potential and superfluid velocity,
while $\rho$ and $J_\varphi$ are the charge density and current in
the dual field theory, respectively. Note that, provided
$\Delta_{-}$ is larger than the unitarity bound, both $\psi_{-}$ and
$\psi_{+}$ can be normalizable and they will be used to define
operators in the dual field theory according to the AdS/CFT
correspondence, $\psi_{-}=\langle O_{-}\rangle$, $\psi_{+}=\langle
O_{+}\rangle$, respectively. We can impose boundary conditions that
either $\psi_{-}$ or $\psi_{+}$ vanishes
\cite{HartnollPRL101,HartnollJHEP12}.

Interestingly, from Eq. (\ref{EqMotionr})  we can get the useful
scaling symmetries
\begin{eqnarray}
r\rightarrow\lambda r\,,\hspace{0.5cm}(t, \varphi, x,
y)\rightarrow\frac{1}{\lambda}(t, \varphi, x,
y)\,,\hspace{0.5cm}(q,\psi)\rightarrow
(q,\psi)\,,\hspace{0.5cm}(A_{t},A_{\varphi})\rightarrow\lambda(A_{t},A_{\varphi})\,,\hspace{0.5cm}
\label{SLsymmetry-1}
\end{eqnarray}
where $\lambda$ is a real positive number. Using these symmetries,
we can obtain the transformation of the relevant quantities
\begin{eqnarray}
(\mu,S_\varphi)\rightarrow\lambda(\mu,S_\varphi)\,,\hspace{0.5cm}(\rho,J_\varphi)\rightarrow\lambda^{3}(\rho,J_\varphi)\,,
\hspace{0.5cm}\psi_{i}\rightarrow\lambda^{\Delta_{i}}\psi_{i}\,,
\label{SLsymmetry-2}
\end{eqnarray}
with $i=+$ or $i=-$. We can use them to set $q=1$ and $r_{s}=1$ when
performing numerical calculations and check the analytical
expressions in this section.

Applying the S-L method to analytically study the properties of the
holographic s-wave model of superfluidity in AdS soliton background,
we will introduce a new variable $z=r_{s}/r$ and rewrite Eq.
(\ref{EqMotionr}) into
\begin{eqnarray}\label{SWEqMotionPsiz}
\psi^{\prime\prime}+\left(\frac{f^\prime}{f}-\frac{1}{z}\right)\psi^\prime
+\left[\frac{1}{z^{2}f}\left(\frac{qA_{t}}{r_{s}}\right)^{2}-\frac{1}{z^{4}f^{2}}\left(\frac{qA_{\varphi}}{r_{s}}\right)^{2}
-\frac{m^{2}}{z^{4}f}\right]\psi=0,
\end{eqnarray}
\begin{eqnarray}\label{SWEqMotionAtz}
A_t^{\prime\prime}+\left(\frac{1}{z}+\frac{f^\prime}{f}\right)A^\prime_{t}-\frac{2q^2\psi^2}{z^{4}f}A_t=0,
\end{eqnarray}
\begin{eqnarray}\label{SWEqMotionAphiz}
A_\varphi^{\prime\prime}-\frac{1}{z}A^\prime_{\varphi}-\frac{2q^2\psi^2}{z^{4}f}A_\varphi=0,
\end{eqnarray}
with $f=(1-z^{4})/z^{2}$. Here and hereafter in this section the
prime denotes the derivative with respect to $z$.

\subsection{Critical chemical potential}

It has been shown numerically that
\cite{Nishioka-Ryu-Takayanagi,HorowitzWay,PanWPOP}, adding the
chemical potential to the AdS soliton, the solution is unstable to
develop a hair for the chemical potential bigger than a critical
value, i.e., $\mu>\mu_{c}$. For lower chemical potential
$\mu<\mu_{c}$, the scalar field is zero and it can be interpreted as
the insulator phase since in this model the normal phase is
described by an AdS soliton where the system exhibits a mass gap.
Therefore, there is a phase transition when $\mu\rightarrow\mu_{c}$
and the AdS soliton reaches the superconductor (or superfluid) phase
for larger $\mu$.

Before going further, we would like to discuss the phase transition
between the AdS soliton and AdS black holes at high chemical
potential without the scalar (or vector) field since it is very
important for us to understand the phase structure of the
holographic dual model in the backgrounds of AdS soliton
\cite{Nishioka-Ryu-Takayanagi,HorowitzWay}. Considering that the
Gibbs Euclidean action of AdS soliton coincides with that of the AdS
charged black hole in the grand canonical ensemble, we find that the
phase boundary between the AdS black hole and the AdS soliton at
zero temperature will be at a chemical potential
$\mu_{d}=2^{1/2}3^{1/4}\simeq1.861$ assuming $r_{s}=1$, which has
been discussed in Refs. \cite{Nishioka-Ryu-Takayanagi,HorowitzWay}.
Obviously, the AdS soliton solution should be replaced with the AdS
black hole at $\mu=\mu_{c}$ and the superconductor (or superfluid)
phase transition gets unphysical if $\mu_{c}>\mu_{d}$. Employing the
analysis of the string theory embedding found in
\cite{GubserPRL2009}, the authors of \cite{Nishioka-Ryu-Takayanagi}
avoided this problem in an explicit string theory setup. In the
following discussion, we will accept this way if we were in a
similar situation.

\begin{figure}[ht]
\includegraphics[scale=0.65]{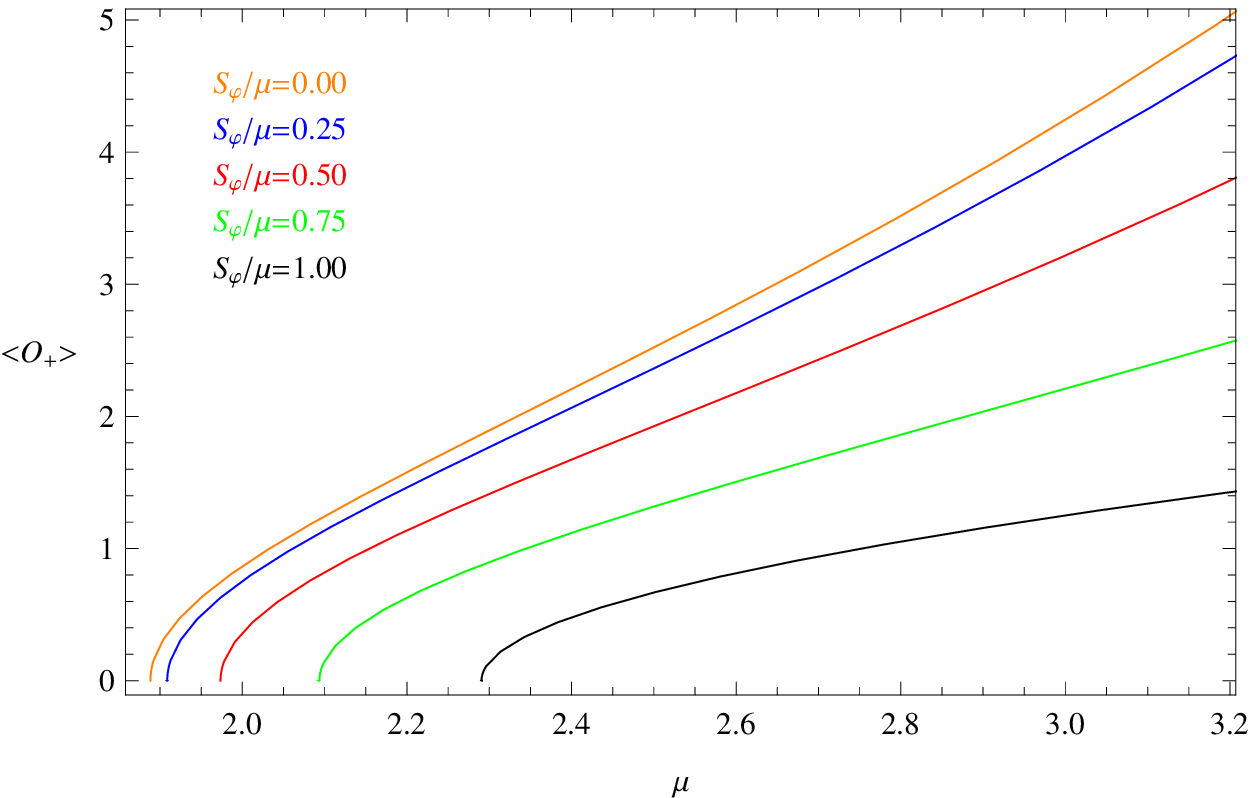}\hspace{0.2cm}%
\includegraphics[scale=0.63]{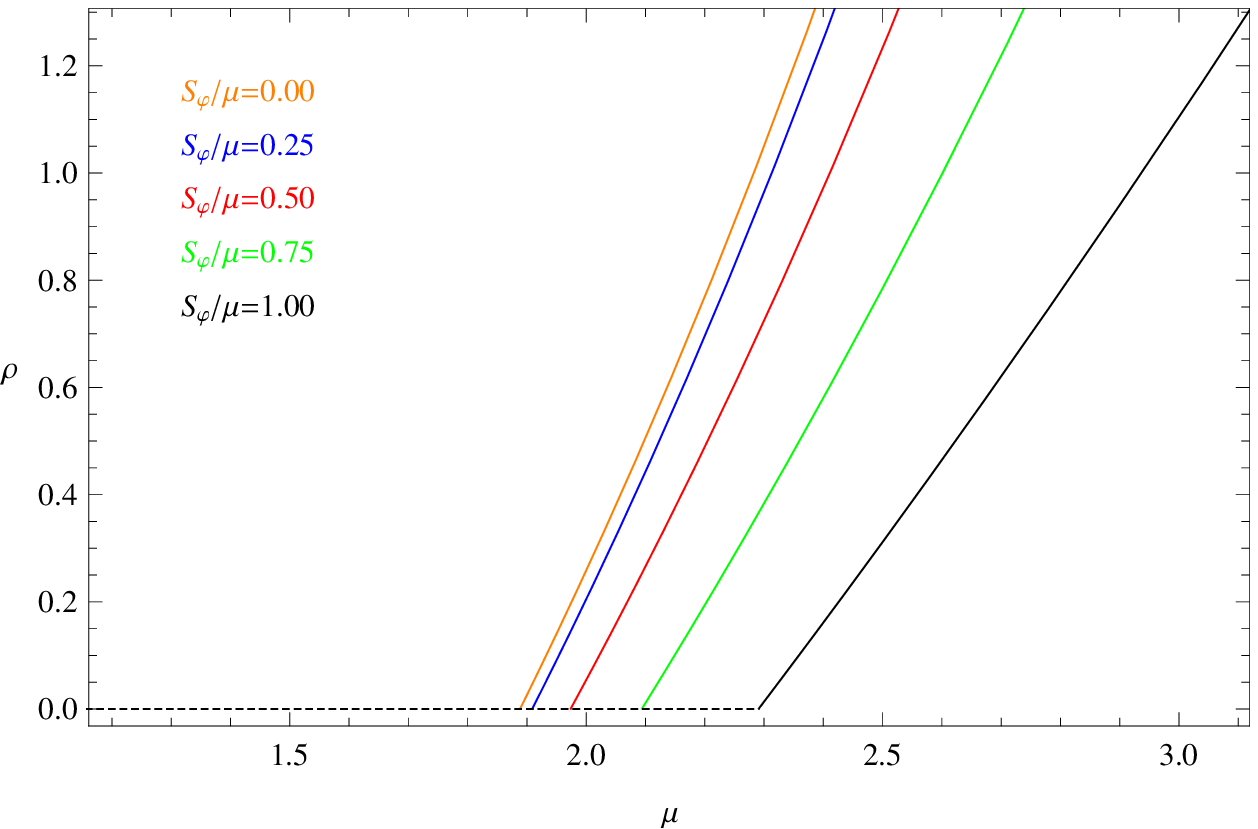}\\ \vspace{0.0cm}
\caption{\label{SWCondRLZheng} (Color online) The condensate of the
operator $\langle O_{+}\rangle$ and charge density $\rho$ with
respect to the chemical potential $\mu$ for different values of the
dimensionless parameter $k=S_{\varphi}/\mu$ in the holographic
s-wave model of superfluidity by using the numerical shooting
method. In each panel, the five lines from left to right correspond
to increasing $S_{\varphi}/\mu$, i.e., $S_{\varphi}/\mu=0.00$
(orange), $0.25$ (blue), $0.50$ (red), $0.75$ (green) and $1.00$
(black) respectively. We choose $m^{2}=-15/4$ and scale $q=1$ and
$r_{s}=1$ in the numerical computation.}
\end{figure}

\begin{figure}[ht]
\includegraphics[scale=0.65]{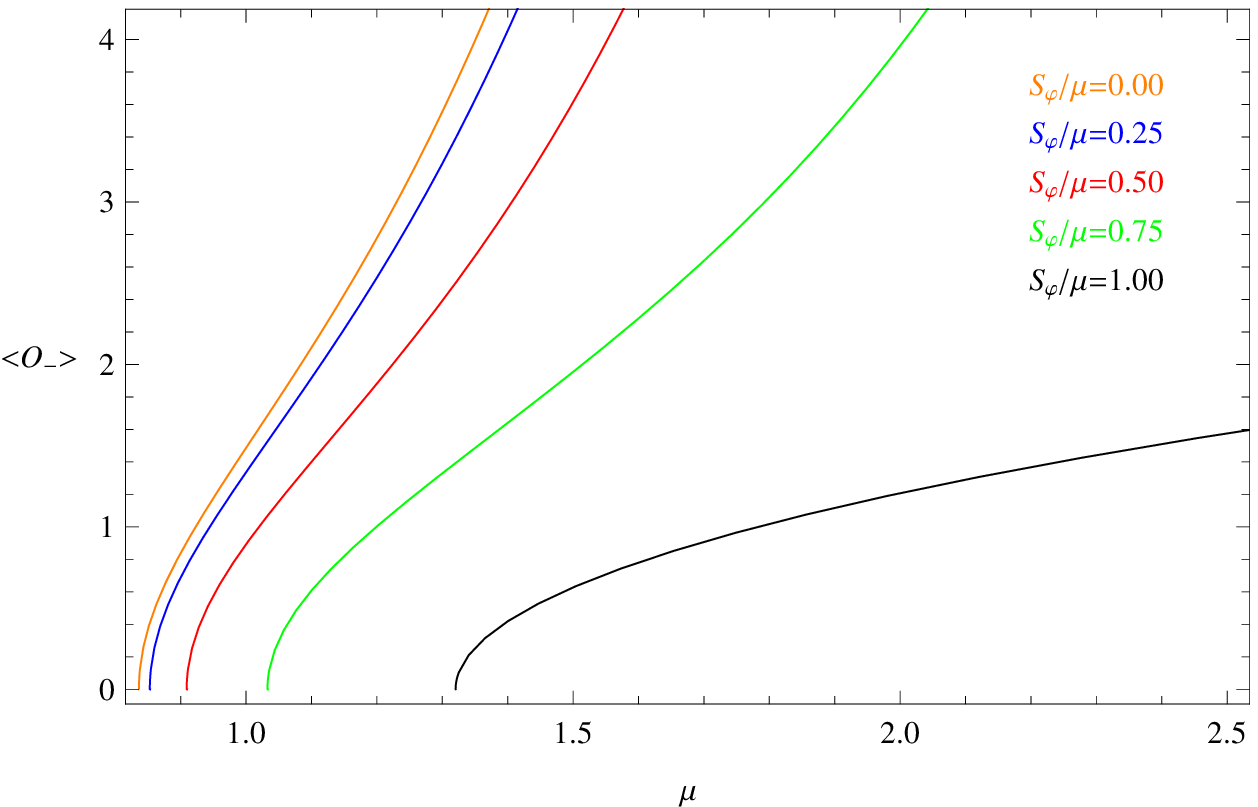}\hspace{0.2cm}%
\includegraphics[scale=0.63]{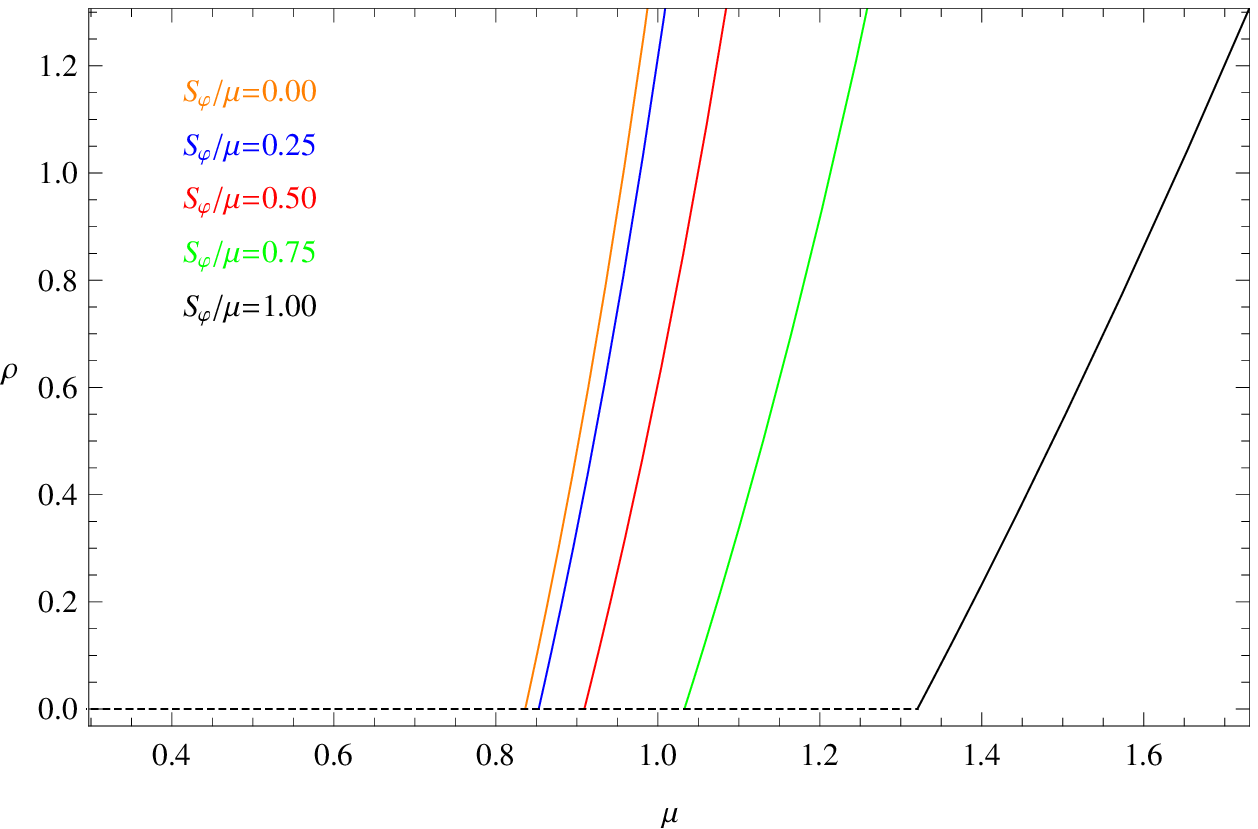}\\ \vspace{0.0cm}
\caption{\label{SWCondRLFu} (Color online)  The condensate of the
operator $\langle O_{-}\rangle$ and charge density $\rho$ with
respect to the chemical potential $\mu$ for different values of the
dimensionless parameter $k=S_{\varphi}/\mu$ in the holographic
s-wave model of superfluidity by using the numerical shooting
method. In each panel, the five lines from left to right correspond
to increasing $S_{\varphi}/\mu$, i.e., $S_{\varphi}/\mu=0.00$
(orange), $0.25$ (blue), $0.50$ (red), $0.75$ (green) and $1.00$
(black) respectively. We choose $m^{2}=-15/4$ and scale $q=1$ and
$r_{s}=1$ in the numerical computation.}
\end{figure}

At the critical chemical potential $\mu_{c}$, the scalar field
$\psi=0$. Thus, below the critical point Eq. (\ref{SWEqMotionAtz})
reduces to
\begin{eqnarray}\label{SWEqMotionAtzCritical}
A_t^{\prime\prime}+\left(\frac{1}{z}+\frac{f^\prime}{f}\right)A^\prime_{t}=0,
\end{eqnarray}
which leads to a general solution
\begin{eqnarray}
A_t=\mu+c_{1}\ln\left(\frac{1+z^2}{1-z^{2}}\right),
\label{SWEqMotionAtzSolution}
\end{eqnarray}
where $c_{1}$ is an integration constant. Obviously, the second term
is divergent at the tip $z=1$ if $c_{1}\neq0$. Considering the
Neumann-like boundary condition (\ref{TipCondition}) for the gauge
field $A_{t}$ at the tip $z=1$, we have to set $c_{1}=0$ to keep
$A_{t}$ finite, i.e., in this case $A_{t}$ will be a constant. Thus,
we can get the physical solution $A_{t}(z)=\mu$ to Eq.
(\ref{SWEqMotionAtzCritical}) if $\mu<\mu_{c}$. This is consistent
with the numerical results in Figs. \ref{SWCondRLZheng} and
\ref{SWCondRLFu} which plot the condensates of the operator $\langle
O_{i}\rangle=\psi_{i}$ and charge density $\rho$ with respect to the
chemical potential $\mu$ for different values of the dimensionless
parameter $k=S_{\varphi}/\mu$.

Similarly, from Eq. (\ref{SWEqMotionAphiz}) we have
\begin{eqnarray}\label{SWEqMAphizCritical}
A_\varphi^{\prime\prime}-\frac{1}{z}A^\prime_{\varphi}=0,
\end{eqnarray}
which results in a solution
\begin{eqnarray} \label{SWEqMAphizSolution}
A_\varphi=S_\varphi(1-z^{2}),
\end{eqnarray}
which is consistent with the boundary condition $A_\varphi(1)=0$
given in (\ref{TipCondition}).

As $\mu\rightarrow\mu_{c}$ from below the critical point, the scalar
field equation (\ref{SWEqMotionPsiz}) becomes
\begin{eqnarray}\label{SWPsizCriticalMotion}
\psi^{\prime\prime}+\left(\frac{f^\prime}{f}-\frac{1}{z}\right)\psi^\prime
+\left[\frac{1}{z^{2}f}\left(\frac{q\mu}{r_{s}}\right)^{2}-\frac{(1-z^{2})^{2}}{z^{4}f^{2}}\left(\frac{qS_\varphi}{r_{s}}\right)^{2}
-\frac{m^{2}}{z^{4}f}\right]\psi=0.
\end{eqnarray}
With the boundary condition (\ref{InfinityCondition}), we assume
$\psi$ takes the form
\begin{eqnarray}\label{SWaveFz}
\psi(z)\sim \frac{\langle
O_{i}\rangle}{r_{s}^{\Delta_{i}}}z^{\Delta_{i}} F(z),
\end{eqnarray}
where the trial function $F(z)$ obeys the boundary conditions
$F(0)=1$ and $F'(0)=0$. From Eq. (\ref{SWPsizCriticalMotion}), we
arrive at
\begin{eqnarray}\label{SWaveFzmotion}
(TF^{\prime})^{\prime}+T\left[U+V\left(\frac{q\mu}{r_{s}}\right)^{2}-W\left(\frac{qS_\varphi}{r_{s}}\right)^{2}\right]F=0,
\end{eqnarray}
where we have defined
\begin{eqnarray}\label{SWaveTUVW}
T=z^{2\Delta_{i}-1}f,~~
U=\frac{\Delta_{i}}{z}\left(\frac{\Delta_{i}-2}{z}+\frac{f^\prime}{f}\right)-\frac{m^2}{z^{4}f},~~
V=\frac{1}{z^{2}f},~~ W=\frac{(1-z^{2})^{2}}{z^{4}f^{2}}.
\end{eqnarray}
According to the Sturm-Liouville eigenvalue problem
\cite{Gelfand-Fomin}, the minimum eigenvalue of $\Lambda=q\mu/r_{s}$
can be obtained from variation of the following functional
\begin{eqnarray}\label{SWSLEigenvalue}
\Lambda^{2}=\left(\frac{q\mu}{r_{s}}\right)^{2}=\frac{\int^{1}_{0}T\left(F'^{2}-UF^{2}\right)dz}{\int^{1}_{0}T(V-k^{2}W)F^{2}dz},
\end{eqnarray}
where we will assume the trial function to be $F(z)=1-az^{2}$ with a
constant $a$. When $k=0$, Eq. (\ref{SWSLEigenvalue}) reduces to the
case considered in \cite{Cai-Li-Zhang} for the holographic s-wave
insulator/superconductor phase transition, where the spatial
component $A_{\varphi}$ has been turned off.

For different values of $k$ and $m^{2}$ with the fixed operator
$\langle O_{+}\rangle$ or $\langle O_{-}\rangle$, we can obtain the
minimum eigenvalue of $\Lambda^{2}$ and the corresponding value of
$a$. As an example, we have $\Lambda_{min}^{2}=3.650$ and $a=0.3214$
for $k=0.25$ with $m^{2}=-15/4$, which gives the critical chemical
potential $\Lambda_{c}=\Lambda_{min}=1.911$ for the operator
$\langle O_{+}\rangle$. In Table \ref{SWaveTable}, we present the
critical chemical potential $\Lambda_{c}=q\mu_{c}/r_{s}$ for chosen
$k$ with fixed mass of the scalar field by $m^{2}=-15/4$ in the
holographic s-wave superfluid model. Obviously, the agreement of the
analytical results derived from the S-L method with the numerical
calculation shown in Table \ref{SWaveTable} is impressive.

\begin{table}[ht]
\caption{\label{SWaveTable} The critical chemical potential
$\Lambda_{c}=q\mu_{c}/r_{s}$ obtained by the analytical S-L method
(left column) and from numerical calculation (right column) with the
chosen values of $k=S_{\varphi}/\mu$ for the scalar operators
$<\mathcal{O}_{-}>$ and $<\mathcal{O}_{+}>$ in the holographic
s-wave superfluid model. Here we fix the mass of the scalar field by
$m^{2}=-15/4$.}
\begin{tabular}{c c c}
         \hline
~ & $<\mathcal{O}_{-}>$ & $<\mathcal{O}_{+}>$
        \\
        \hline
~~~~$k=0.00$~~~~~~~~&~~~~~~~~$0.8368$~~~~~~~~$0.8362$~~~~~~~~&~~~~~~~~$1.890$~~~~~~~~$1.888$~~~~~
          \\
~~~~$k=0.25$~~~~~~~~&~~~~~~~~$0.8534$~~~~~~~~$0.8528$~~~~~~~~&~~~~~~~~$1.911$~~~~~~~~$1.909$~~~~~
          \\
~~~~$k=0.50$~~~~~~~~&~~~~~~~~$0.9096$~~~~~~~~$0.9092$~~~~~~~~&~~~~~~~~$1.975$~~~~~~~~$1.973$~~~~~
          \\
~~~~$k=0.75$~~~~~~~~&~~~~~~~~$1.032(7)$~~~~~~~~$1.032(8)$~~~~~~~~&~~~~~~~~$2.094$~~~~~~~~$2.067$~~~~~
          \\
~~~~$k=1.00$~~~~~~~~&~~~~~~~~$1.320(5)$~~~~~~~~$1.320(3)$~~~~~~~~&~~~~~~~~$2.291$~~~~~~~~$2.290$~~~~~
          \\
        \hline
\end{tabular}
\end{table}

We see that, from Table \ref{SWaveTable} and Figs.
\ref{SWCondRLZheng} and \ref{SWCondRLFu}, the critical chemical
potential $\Lambda_{c}=q\mu_{c}/r_{s}$ increases as the
dimensionless parameter $k=S_{\varphi}/\mu$ increases for the fixed
mass of the scalar field, i.e., the critical chemical potential
becomes larger with the increase of the superfluid velocity, which
indicates that the spatial component of the gauge field to modeling
the superfluid hinders the phase transition. This result is
consistent with the observation obtained from the effective mass of
the scalar field in Eq. (\ref{EffectiveMass}), which implies that
the increasing $A_\varphi$ will hinder the s-wave superfluid phase
transition.

\subsection{Critical phenomena}
Now we are in a position to study the critical phenomena of this
holographic s-wave superfluid system. Considering that the
condensation of the scalar operator $\langle O_{i}\rangle$ is so
small near the critical point, we can expand $A_{t}(z)$ in $\langle
O_{i}\rangle$ as
\begin{eqnarray}\label{SWAtz}
A_{t}(z)\sim\mu_{c}+\langle O_{i}\rangle\chi(z)+\cdots,
\end{eqnarray}
where we have introduced the boundary condition $\chi(1)=0$ at the
tip. Defining a function $\xi(z)$ as
\begin{eqnarray}\label{SWXiz}
\chi(z)=\frac{2q^{2}\mu_{c}}{r_{s}^{2\Delta_{i}}}\langle
O_{i}\rangle\xi(z),
\end{eqnarray}
we obtain the equation of motion for $\xi(z)$
\begin{eqnarray}\label{SWXiEoM}
(Q\xi')'-z^{2\Delta_{i}-3}F^2=0,
\end{eqnarray}
with
\begin{eqnarray}\label{SWQz}
Q(z)=zf(z).
\end{eqnarray}

According to the asymptotic behavior in Eq.
(\ref{InfinityCondition}) and Eq. (\ref{SWXiz}), we will expand
$A_{t}$ when $z\rightarrow0$ as
\begin{eqnarray}\label{SWPhiExpand}
A_{t}(z)\simeq\mu-\frac{\rho}{r_{s}^{2}}z^2\simeq\mu_c
+2\mu_{c}\left(\frac{q\langle
O_{i}\rangle}{r_{s}^{\Delta_{i}}}\right)^{2}\left[\xi(0)+\xi^\prime(0)z+\frac{1}{2}\xi^{\prime\prime}(0)z^2+\cdot\cdot\cdot\right].
\end{eqnarray}
From the coefficients of the $z^0$ term in both sides of the above
formula, we have
\begin{eqnarray}\label{SWOxExp}
\frac{q\langle
O_{i}\rangle}{r_{s}^{\Delta_{i}}}=\frac{1}{\left[2\mu_c\xi(0)\right]^{\frac{1}{2}}}\left(\mu-\mu_c\right)^{\frac{1}{2}},
\end{eqnarray}
with
\begin{eqnarray}
\xi(0)=c_2-\int^1_0\frac{1}{Q(z)}\left[c_3+\int^z_1
x^{2\Delta_{i}-3}F(x)^2dx\right]dz,
\end{eqnarray}
where $c_{2}$ and $c_{3}$ are the integration constants which can be
determined by the boundary condition of $\chi(z)$. For example, for
the case of $k=0.25$ with $m^{2}=-15/4$, we have $\langle
O_{+}\rangle\approx1.776(\mu-\mu_{c})^{1/2}$ when $a=0.3214$ (we
have scaled $q=1$ and $r_{s}=1$ for simplicity), which is in good
agreement with the numerical result shown in the left panel of Fig.
\ref{SWCondRLZheng}. Note that our expression (\ref{SWOxExp}) is
valid for all cases considered here, so near the critical point,
both of the scalar operators $\langle O_{+}\rangle$ and $\langle
O_{-}\rangle$ satisfy $\langle O_{i}\rangle\sim(\mu-\mu_{c})^{1/2}$.
This analytical result shows that the phase transition of the
holographic s-wave superfluid model belongs to the second order and
the critical exponent of the system takes the mean-field value
$1/2$, which can be used to back up the numerical findings as shown
in Figs. \ref{SWCondRLZheng} and \ref{SWCondRLFu}.

Comparing the coefficients of the $z^1$ term in Eq.
(\ref{SWPhiExpand}), we observe that $\xi^\prime(0)\rightarrow 0$,
which agrees well with the following relation by making integration
of both sides of Eq. (\ref{SWXiEoM})
\begin{eqnarray}
\left[\frac{\xi'(z)}{z}\right]\bigg|_{z\rightarrow
0}=-\int_{0}^{1}z^{2\Delta_{i}-3}F^2dz.
\end{eqnarray}

Considering the coefficients of the $z^2$ term in Eq.
(\ref{SWPhiExpand}), we get
\begin{eqnarray}\label{SWRhoExp}
\frac{\rho}{r_{s}^{2}}=-\left(\frac{q\langle
O_{i}\rangle}{r_{s}^{\Delta_{i}}}\right)^{2}\mu_{c}\xi^{\prime\prime}(0)=\Gamma(k,m)(\mu-\mu_{c}),
\end{eqnarray}
with a prefactor
\begin{eqnarray}
\Gamma(k,m)=\frac{1}{2\xi(0)}\int_{0}^{1}z^{2\Delta_{i}-3}F^2dz,
\end{eqnarray}
which is a function of the parameter $k$ and scalar field mass
$m^{2}$. For the case of $k=0.25$ and $m^{2}=-15/4$ with the
operator $\langle O_{+}\rangle$, as an example, we can obtain
$\rho=1.323\left(\mu-\mu_c\right)$ when $a=0.3214$ (we have scaled
$q=1$ and $r_{s}=1$ for simplicity), which is consistent with the
result given in the right panel of Fig. \ref{SWCondRLZheng}.
Obviously, the parameter $k$ and mass of the scalar field $m^{2}$
will not change the linear relation between the charge density and
chemical potential near $\mu_{c}$, i.e., $\rho\sim(\mu-\mu_{c})$,
which is in good agreement with the numerical results plotted in
Figs. \ref{SWCondRLZheng} and \ref{SWCondRLFu}.

On the other hand, near the critical point Eq.
(\ref{SWEqMotionAphiz}) becomes
\begin{eqnarray}\label{SWEMAphizCri}
A_\varphi^{\prime\prime}-\frac{1}{z}A^\prime_{\varphi}-\frac{2S_{\varphi}(1-z^{2})}{z^{4}f}
\left(\frac{q\langle O_{i}\rangle
z^{\Delta_{i}}F}{r_{s}^{\Delta_{i}}}\right)^{2}=0.
\end{eqnarray}
Thus, we finally arrive at
\begin{eqnarray}\label{SWEMAphizSolution}
A_\varphi=S_{\varphi}(1-z^{2})+S_{\varphi}\left(\frac{q\langle
O_{i}\rangle}{r_{s}^{\Delta_{i}}}\right)^{2}\int
z\left[\int\frac{2x^{2\Delta_{i}-5}(1-x^{2})F(x)^{2}}{f(x)}dx\right]dz,
\end{eqnarray}
which obeys the boundary condition $A_\varphi(1)=0$ presented in
(\ref{TipCondition}) at the critical point. For example, for the
case of $k=0.25$ with $m^{2}=-15/4$, we obtain
$A_\varphi=S_{\varphi}[(1-z^{2})-0.07382\langle
O_{+}\rangle^{2}z^{2}+\cdot\cdot\cdot$] when $a=0.3214$ (we have
scaled $q=1$ and $r_{s}=1$ for simplicity) for the operator $\langle
O_{+}\rangle$. Obviously, Eq. (\ref{SWEMAphizSolution}) is
consistent with the behavior of $A_{\varphi}$ in Eq.
(\ref{SWEqMAphizSolution}) at the critical point.

\section{Holographic p-wave superfluid model}

Since the S-L method is effective to obtain the properties of the
holographic s-wave model of superfluidity in the AdS soliton
background, we will use it to investigate analytically the
holographic p-wave model of superfluidity in the AdS soliton
background which has not been constructed as far as we know.

Considering the Maxwell complex vector field model which was first
proposed in \cite{CaiPWave-1,CaiPWave-2}, we will build the
holographic p-wave model of superfluidity in the AdS soliton
background via the action
\begin{eqnarray}\label{PWaveAtion}
S=\frac{1}{16\pi G}\int
d^{5}x\sqrt{-g}\left(-\frac{1}{4}F_{\mu\nu}F^{\mu\nu}-\frac{1}{2}\rho_{\mu\nu}^{\dag}\rho^{\mu\nu}-m^2\rho_{\mu}^{\dag}\rho^{\mu}+i
q\gamma\rho_{\mu}\rho_{\nu}^{\dag}F^{\mu\nu} \right),
\end{eqnarray}
where the tensor $\rho_{\mu\nu}$ is defined by
$\rho_{\mu\nu}=D_\mu\rho_\nu-D_\nu\rho_\mu$ with the covariant
derivative $D_\mu=\nabla_\mu-iqA_\mu$, $q$ and $m$ are the charge
and mass of the vector field $\rho_\mu$, respectively. Since we
consider the case without external magnetic field in this work, the
parameter $\gamma$, which describes the interaction between the
vector field $\rho_\mu$ and the gauge field $A_\mu$, will not play
any role.

As in Refs. \cite{PWavSuperfluidA,LifshitzSuperfluid}, we take the
same ansatz for the gauge field $A_\mu$ just as in Eq.
(\ref{Ansatz}) and assume the condensate to pick out the $x$
direction as special
\begin{eqnarray}\label{PWaveAnsatz}
\rho_\mu dx^{\mu}=\rho_{x}(r)dx,
\end{eqnarray}
where we can set $\rho_{x}(r)$ to be real by using the $U(1)$ gauge
symmetry. Thus, in the soliton background (\ref{SchSoliton}), we can
obtain the equations of motion for the holographic p-wave superfluid
model
\begin{eqnarray}\label{PWaveEqMotionr}
&&\rho_{x}^{\prime\prime}+\left(\frac{1}{r}+\frac{f^\prime}{f}\right)\rho_{x}^{\prime}
-\frac{1}{f}\left(m^2+\frac{q^2A^2_\varphi}{f}-\frac{q^2A_t^2}{r^2}\right)\rho_{x}=0,\nonumber \\
&&A_t^{\prime\prime}+\left(\frac{1}{r}+\frac{f^\prime}{f}\right)A_t^{\prime}-\frac{2q^2\rho_{x}^2}{r^{2}f}A_t=0,\nonumber \\
&&A_\varphi^{\prime\prime}+\frac{3}{r}A_\varphi^{\prime}-\frac{2q^2\rho_{x}^2}{r^{2}f}A_\varphi=0,
\end{eqnarray}
where the prime denotes the derivative with respect to $r$.
Obviously, we find that the effective mass of the vector field has
the same expression just as in (\ref{EffectiveMass}), which means
that the increasing $m^2$ and $A_\varphi$ or decreasing $A_t$ will
hinder the p-wave superfluid phase transition.

Analyzing the boundary conditions of the matter fields, we observe
that $A_{t}$ and $A_{\varphi}$ have the same boundary conditions
just as Eq. (\ref{TipCondition}) for the tip $r=r_{s}$ and Eq.
(\ref{InfinityCondition}) for the boundary $r\rightarrow\infty$. But
for the vector field $\rho_{x}$, we find that at the tip
\begin{eqnarray}\label{PWTipCondition}
\rho_{x}=\tilde{\rho}_{x0}+\tilde{\rho}_{x1}(r-r_{s})+\tilde{\rho}_{x2}(r-r_{s})^{2}+\cdots\,,
\end{eqnarray}
with the integration constant $\tilde{\rho}_{xi}$
($i=0,1,2,\cdots$), and at the asymptotic AdS boundary
\begin{eqnarray}\label{PWInfinityCondition}
\rho_{x}=\frac{\rho_{x-}}{r^{2-\Delta}}+\frac{\rho_{x+}}{r^{\Delta}},
\end{eqnarray}
with the characteristic exponent $\Delta=1+\sqrt{1+m^2}$. According
to the AdS/CFT correspondence, $\rho_{x-}$ and $\rho_{x+}$ are
interpreted as the source and the vacuum expectation value of the
vector operator $\langle O_{x}\rangle$ in the dual field theory
respectively. Since we require that the condensate appears
spontaneously, we will impose boundary condition $\rho_{x-}=0$ in
this work.

From Eq. (\ref{PWaveEqMotionr}), we can also have the useful scaling
symmetries
\begin{eqnarray}
r\rightarrow\lambda r\,,\hspace{0.5cm}(t, \varphi, x,
y)\rightarrow\frac{1}{\lambda}(t, \varphi, x,
y)\,,\hspace{0.5cm}q\rightarrow
q\,,\hspace{0.5cm}(\rho_{x},A_{t},A_{\varphi})\rightarrow\lambda(\rho_{x},A_{t},A_{\varphi})\,,
\label{PWSLsymmetry-1}
\end{eqnarray}
and the transformation of the relevant quantities
\begin{eqnarray}
(\mu,S_\varphi)\rightarrow\lambda(\mu,S_\varphi)\,,\hspace{0.5cm}
(\rho,J_\varphi)\rightarrow\lambda^{3}(\rho,J_\varphi)\,,\hspace{0.5cm}
\rho_{x+}\rightarrow\lambda^{1+\Delta}\rho_{x+}\,,
\label{PWSLsymmetry-2}
\end{eqnarray}
which can be used to set $q=1$ and $r_{s}=1$ when performing
numerical calculations and check the analytical expressions in this
section.

For convenience in the following discussion, we will change the
coordinate $z=r_{s}/r$ and convert Eq. (\ref{PWaveEqMotionr}) to be
\begin{eqnarray}\label{PWEqMotionRhoz}
\rho_{x}^{\prime\prime}+\left(\frac{1}{z}+\frac{f^\prime}{f}\right)\rho_{x}^\prime
+\left[\frac{1}{z^{2}f}\left(\frac{qA_{t}}{r_{s}}\right)^{2}-\frac{1}{z^{4}f^{2}}\left(\frac{qA_{\varphi}}{r_{s}}\right)^{2}
-\frac{m^{2}}{z^{4}f}\right]\rho_{x}=0,
\end{eqnarray}
\begin{eqnarray}\label{PWEqMotionAtz}
A_t^{\prime\prime}+\left(\frac{1}{z}+\frac{f^\prime}{f}\right)A^\prime_{t}-\frac{2}{z^{2}f}\left(\frac{q\rho_{x}}{r_{s}}\right)^{2}A_t=0,
\end{eqnarray}
\begin{eqnarray}\label{PWEqMotionAphiz}
A_\varphi^{\prime\prime}-\frac{1}{z}A^\prime_{\varphi}-\frac{2}{z^{2}f}\left(\frac{q\rho_{x}}{r_{s}}\right)^{2}A_\varphi=0.
\end{eqnarray}
Here and hereafter in this section the prime denotes the derivative
with respect to $z$.

\subsection{Critical chemical potential}

\begin{figure}[ht]
\includegraphics[scale=0.65]{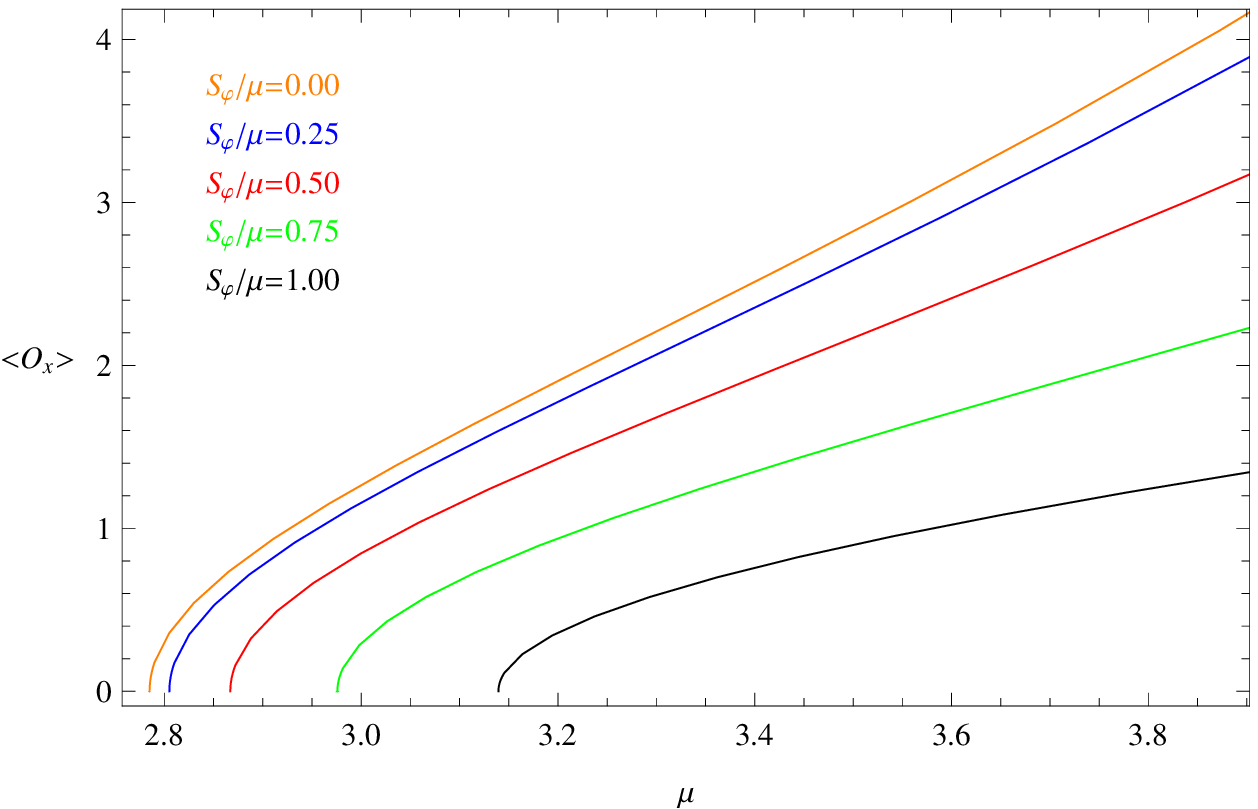}\hspace{0.2cm}%
\includegraphics[scale=0.63]{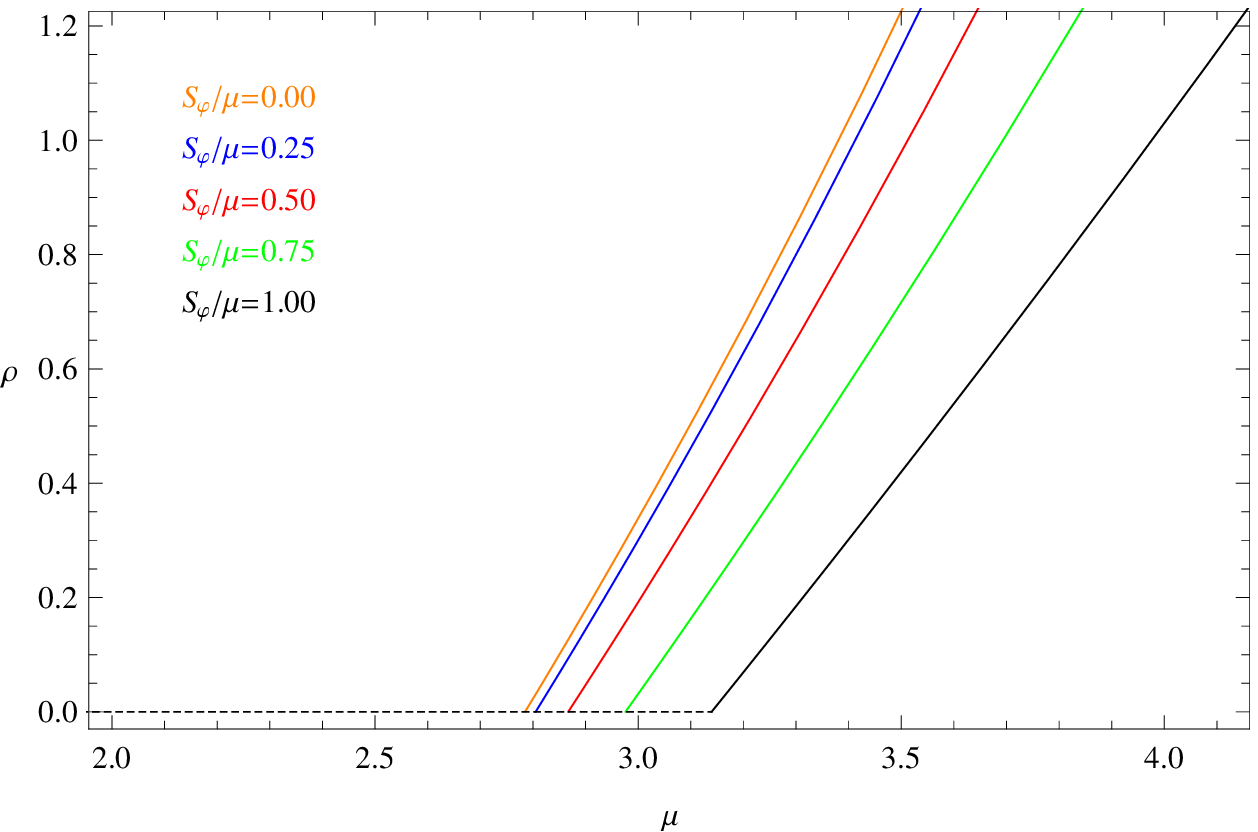}\\ \vspace{0.0cm}
\caption{\label{PWCondRL} (Color online)  The condensate of the
operator $\langle O_{x}\rangle$ and charge density $\rho$ with
respect to the chemical potential $\mu$ for different values of
$k=S_{\varphi}/\mu$ in the holographic p-wave model of superfluidity
by using the numerical shooting method. In each panel, the five
lines from left to right correspond to increasing $S_{\varphi}/\mu$,
i.e., $S_{\varphi}/\mu=0.00$ (orange), $0.25$ (blue), $0.50$ (red),
$0.75$ (green) and $1.00$ (black) respectively. We choose
$m^{2}=5/4$ and scale $q=1$ and $r_{s}=1$ in the numerical
computation.}
\end{figure}

Similar to the analysis for the holographic s-wave model of
superfluidity, if $\mu\leq\mu_{c}$, the vector field $\rho_x$ is
nearly zero, i.e., $\rho_x\simeq0$. Thus, we can obtain the physical
solutions $A_{t}(z)=\mu$ to Eq. (\ref{PWEqMotionAtz}) and
$A_\varphi=S_\varphi(1-z^{2})$ to Eq. (\ref{PWEqMotionAphiz}) when
$\mu<\mu_{c}$, which are the same forms just as in the holographic
s-wave superfluid model. This analytical result is consistent with
the numerical findings in Fig. \ref{PWCondRL} which plots the
condensate of the operator $\langle O_{x}\rangle=\rho_{x+}$ and
charge density $\rho$ with respect to the chemical potential $\mu$
for different values of the dimensionless parameter
$k=S_{\varphi}/\mu$.

As $\mu\rightarrow\mu_{c}$, the vector field equation
(\ref{PWEqMotionRhoz}) will become
\begin{eqnarray}\label{PWRhozCriMotion}
\rho_{x}^{\prime\prime}+\left(\frac{1}{z}+\frac{f^\prime}{f}\right)\rho_{x}^\prime
+\left[\frac{1}{z^{2}f}\left(\frac{q\mu}{r_{s}}\right)^{2}-\frac{(1-z^{2})^{2}}{z^{4}f^{2}}\left(\frac{qS_\varphi}{r_{s}}\right)^{2}
-\frac{m^{2}}{z^{4}f}\right]\rho_{x}=0.
\end{eqnarray}
Defining a trial function $F(z)$ which matches the boundary behavior
(\ref{PWInfinityCondition}) for $\rho_x$ \cite{Siopsis}
\begin{eqnarray}\label{SLFz}
\rho_x(z)\sim \frac{\langle O_{x}\rangle}{r_{s}^{\Delta}}z^{\Delta}
F(z),
\end{eqnarray}
with the boundary conditions $F(0)=1$ and $F'(0)=0$, from Eq.
(\ref{PWRhozCriMotion}) we can get the equation of motion for $F(z)$
\begin{eqnarray}\label{SLFzmotion}
(MF^{\prime})^{\prime}+M\left[P+V\left(\frac{q\mu}{r_{s}}\right)^{2}-W\left(\frac{qS_\varphi}{r_{s}}\right)^{2}\right]F=0,
\end{eqnarray}
with
\begin{eqnarray}\label{PWaveTUVWFu}
M=z^{1+2\Delta}f,~~
P=\frac{\Delta}{z}\left(\frac{\Delta}{z}+\frac{f^\prime}{f}\right)-\frac{m^2}{z^{4}f},
\end{eqnarray}
where $V(z)$ and $W(z)$ have been introduced in (\ref{SWaveTUVW}).
Following the S-L eigenvalue problem \cite{Gelfand-Fomin}, we deduce
the eigenvalue $\Lambda=q\mu/r_{s}$ minimizes the expression
\begin{eqnarray}\label{PWSLEigenvalue}
\Lambda^{2}=\left(\frac{q\mu}{r_{s}}\right)^{2}=\frac{\int^{1}_{0}M\left(F'^{2}-PF^{2}\right)dz}{\int^{1}_{0}M(V-k^{2}W)F^{2}dz},
\end{eqnarray}
where we still assume the trial function to be $F(z)=1-az^{2}$ with
a constant $a$. When the dimensionless parameter $k=0$, Eq.
(\ref{PWSLEigenvalue}) reduces to the case considered in
\cite{ZPJ2015} for the holographic p-wave insulator/superconductor
phase transition, where the spatial component $A_{\varphi}$ has been
turned off.

For different values of $k$ and $m^{2}$, we can get the minimum
eigenvalue of $\Lambda^{2}$ and the corresponding value of $a$, for
example, $\Lambda_{min}^{2}=7.879$ and $a=0.3716$ for $k=0.25$ with
$m^{2}=5/4$, which leads to the critical chemical potential
$\Lambda_{c}=\Lambda_{min}=2.807$. In Table \ref{SLTable}, we
present the critical chemical potential $\Lambda_{c}=q\mu_{c}/r_{s}$
for chosen $k$. In order to compare with numerical results given in
Fig. \ref{PWCondRL}, we fix the mass of the vector field by
$m^{2}=5/4$. Obviously, the analytical results derived from S-L
method are in very good agreement with the numerical computations.

\begin{table}[ht]
\begin{center}
\caption{\label{SLTable} The critical chemical potential
$\Lambda_{c}=q\mu_{c}/r_{s}$ for the vector operator
$\langle{O_{x}}\rangle$ obtained by the analytical S-L method and
numerical shooting method with chosen $k=S_{\varphi}/\mu$ for the
fixed mass of the vector field $m^{2}=5/4$ in the holographic p-wave
superfluid model.}
\begin{tabular}{c c c c c c c c c}
         \hline \hline
$k$ & 0.00 & 0.25 & 0.50 & 0.75 & 1.00
        \\
        \hline
~~~~~Analytical~~~~~&~~~~~~$2.787$~~~~~~&~~~~~~$2.807$~~~~~~&~~~~~~$2.868$~~~~~~
&~~~~~~$2.976(1)$~~~~~~&~~~~~~$3.140$~~~~~
          \\
~~~~~Numerical~~~~~&~~~~~~$2.785$~~~~~~&~~~~~~$2.805$~~~~~~&~~~~~~$2.867$~~~~~~
&~~~~~~$2.975(8)$~~~~~~&~~~~~~$3.139$~~~~~
          \\
        \hline \hline
\end{tabular}
\end{center}
\end{table}

From Table \ref{SLTable}, we observe that, for the fixed mass of the
vector field, the critical chemical potential
$\Lambda_{c}=q\mu_{c}/r_{s}$ becomes larger with the increase of
$k=S_{\varphi}/\mu$, i.e., the critical chemical potential increases
with the increase of the superfluid velocity. The fact implies that
the spatial component of the gauge field to modeling the superfluid
hinders the phase transition, which supports the observation
obtained from the effective mass of the vector field in Eq.
(\ref{PWaveEqMotionr}).

\subsection{Critical phenomena}

Since the condensation of the vector operator $\langle O_{x}\rangle$
is so small when $\mu\rightarrow\mu_{c}$, we can expand $A_{t}(z)$
in small $\langle O_{x}\rangle$ as
\begin{eqnarray}\label{PWEigenvalue}
A_{t}(z)\sim\mu_{c}+\langle O_{x}\rangle\chi(z)+\cdots,
\end{eqnarray}
with the boundary condition $\chi(1)=0$ at the tip. Introducing a
function $\xi(z)$ as
\begin{eqnarray}\label{PWXiz}
\chi(z)=\frac{2q^{2}\mu_{c}}{r_{s}^{2(1+\Delta)}}\langle
O_{x}\rangle\xi(z),
\end{eqnarray}
we get the equation of motion for $\xi(z)$
\begin{eqnarray}\label{PWXiEoM}
(Q\xi')'-z^{2\Delta-1}F^2=0,
\end{eqnarray}
where $Q(z)$ has been defined in (\ref{SWQz}).

Considering the asymptotic behavior of $A_{t}$ and Eq.
(\ref{PWXiz}), near $z\rightarrow0$ we will expand $A_{t}$ as
\begin{eqnarray}\label{PWPhiExpand}
A_{t}(z)\simeq\mu-\frac{\rho}{r_{s}^{2}}z^2\simeq\mu_c
+2\mu_{c}\left(\frac{q\langle
O_{x}\rangle}{r_{s}^{1+\Delta}}\right)^{2}\left[\xi(0)+\xi^\prime(0)z+\frac{1}{2}\xi^{\prime\prime}(0)z^2+\cdot\cdot\cdot\right].
\end{eqnarray}
According to the coefficients of the $z^0$ term in both sides of the
above formula, we obtain
\begin{eqnarray}\label{PWOxExp}
\frac{q\langle
O_{x}\rangle}{r_{s}^{1+\Delta}}=\frac{1}{\left[2\mu_c\xi(0)\right]^{\frac{1}{2}}}\left(\mu-\mu_c\right)^{\frac{1}{2}},
\end{eqnarray}
with
\begin{eqnarray}
\xi(0)=c_2-\int^1_0\frac{1}{Q(z)}\left[c_3+\int^z_1
x^{2\Delta-1}F(x)^2dx\right]dz,
\end{eqnarray}
where $c_{2}$ and $c_{3}$ are the integration constants which can be
determined by the boundary condition of $\chi(z)$. For example, for
the case of $k=0.25$ with $m^{2}=5/4$, we have $\langle
O_{x}\rangle\approx1.818(\mu-\mu_{c})^{1/2}$ when $a=0.3716$ (we
have scaled $q=1$ and $r_{s}=1$ for simplicity), which agrees well
with the numerical result shown in the left panel of Fig.
\ref{PWCondRL}. Obviously, the expression (\ref{PWOxExp}) is valid
for all cases considered here. Since the parameter $k$ and mass of
the vector field $m^{2}$ will not alter Eq. (\ref{PWOxExp}) except
for the prefactor, we can obtain the relation $\langle
O_{x}\rangle\sim\left(\mu-\mu_c\right)^{1/2}$ near the critical
point, which shows that the phase transition of the holographic
p-wave superfluid model belongs to the second order and the critical
exponent of the system takes the mean-field value $1/2$. The
analytic result supports the numerical findings obtained from the
left panel of Fig. \ref{PWCondRL}.

From the coefficients of the $z^1$ term in Eq. (\ref{PWPhiExpand}),
we find that $\xi^\prime(0)\rightarrow 0$, which is consistent with
the following relation by making integration of both sides of Eq.
(\ref{PWXiEoM})
\begin{eqnarray}
\left[\frac{\xi'(z)}{z}\right]\bigg|_{z\rightarrow
0}=-\int_{0}^{1}z^{2\Delta-1}F^2dz.
\end{eqnarray}

Comparing the coefficients of the $z^2$ term in Eq.
(\ref{PWPhiExpand}), we arrive at
\begin{eqnarray}\label{PWRhoExp}
\frac{\rho}{r_{s}^{2}}=-\left(\frac{q\langle
O_{x}\rangle}{r_{s}^{1+\Delta}}\right)^{2}\mu_{c}\xi^{\prime\prime}(0)=\Gamma(k,m)(\mu-\mu_{c}),
\end{eqnarray}
with
\begin{eqnarray}
\Gamma(k,m)=\frac{1}{2\xi(0)}\int_{0}^{1}z^{2\Delta-1}F^2dz,
\end{eqnarray}
which is a function of the parameter $k$ and vector field mass
$m^{2}$. For the case of $k=0.25$ with $m^{2}=5/4$, as an example,
we can obtain $\rho=1.013\left(\mu-\mu_c\right)$ when $a=0.3716$ (we
have scaled $q=1$ and $r_{s}=1$ for simplicity), which is in good
agreement with the result shown in the right panel of Fig.
\ref{PWCondRL}. Note that the parameter $k$ and mass of the vector
field $m^{2}$ will not alter Eq. (\ref{PWRhoExp}), we can obtain the
linear relation between the charge density and chemical potential
near $\mu_{c}$, i.e., $\rho\sim(\mu-\mu_{c})$, which can be used to
back up the numerical result presented in the right panel of Fig.
\ref{PWCondRL}.

Similarly, considering Eq. (\ref{PWEqMotionAphiz}) near the phase
transition point, i.e.,
\begin{eqnarray}\label{PWEMAphizCri}
A_\varphi^{\prime\prime}-\frac{1}{z}A^\prime_{\varphi}-\frac{2S_{\varphi}(1-z^{2})}{z^{2}f}
\left(\frac{q\langle O_{x}\rangle
z^{\Delta}F}{r_{s}^{1+\Delta}}\right)^{2}=0,
\end{eqnarray}
we can solve it and get
\begin{eqnarray}\label{PWEMAphizSolution}
A_\varphi=S_{\varphi}(1-z^{2})+S_{\varphi}\left(\frac{q\langle
O_{x}\rangle}{r_{s}^{1+\Delta}}\right)^{2}\int
z\left[\int\frac{2x^{2\Delta-3}(1-x^{2})F(x)^{2}}{f(x)}dx\right]dz,
\end{eqnarray}
which is consistent with the boundary condition $A_\varphi(1)=0$ at
the critical point. For example, for the case of $k=0.25$ with
$m^{2}=5/4$, we have $A_\varphi=S_{\varphi}[(1-z^{2})-0.02450\langle
O_{x}\rangle^{2}z^{2}+\cdot\cdot\cdot$] when $a=0.3716$ (we have
scaled $q=1$ and $r_{s}=1$ for simplicity), which supports our
numerical computation.

\section{Conclusions}

We have applied the S-L method to study analytically the properties
of the holographic superfluid models in the AdS soliton background
in order to understand the influence of the spatial component of the
gauge field on the superfluid phase transition. By investigating the
s-wave (the scalar field) and p-wave (the vector field) models in
the probe limit, we obtained analytically the critical chemical
potentials which are perfectly in agreement with those obtained from
numerical computations. We observed that the critical chemical
potential increases with the increase of the superfluid velocity,
which indicates that the spatial component of the gauge field
hinders the phase transition. Moreover, we found that in the
superfluid model the S-L method can present us analytical results on
the critical exponent of condensation operator, the relation between
the charge density and chemical potential, and the behavior of the
spatial component of the gauge field near the phase transition
point. In particular, we analytically demonstrated that, different
from the findings as shown in the AdS black hole background where
the spatial component of the gauge field can determine the order of
the superfluid phase transition, in the AdS soliton the first-order
phase transition cannot be brought by the supercurrent, i.e., the
holographic superfluid phase transition always belongs to the second
order and the critical exponent of the system takes the mean-field
value $1/2$ in both s-wave and p-wave models. The analytical results
can be used to back up the numerical findings in both holographic
s-wave \cite{KuangLiuWang} and p-wave superfluid models in the AdS
soliton background. Since the superfluid velocity provides richer
physics in the superfluid phase transition in the AdS black hole
background \cite{BasuMukherjeeShieh,HerzogKovtunSon}, it would be of
interest to generalize our study to the AdS black hole configuration
and analytically discuss the effect of the spatial component of the
gauge field on the system. We will leave it for further study.

\begin{acknowledgments}

We thank Professor Rong-Gen Cai for his helpful discussions and
suggestions. This work was supported by the National Natural Science
Foundation of China under Grant Nos. 11275066 and 11475061; Hunan
Provincial Natural Science Foundation of China under Grant Nos.
12JJ4007 and 11JJ7001; and FAPESP No. 2013/26173-9.

\end{acknowledgments}

\end{document}